\pdfoutput=1
\documentclass[aps,twocolumn,superscriptaddress,letterpaper]{revtex4}
\usepackage{graphicx,amssymb,bm,natbib,amsfonts,amsmath}
\usepackage{hyperref}

\begin{document}

\title{Instability of two dimensional graphene: Breaking sp$^2$ bonds with soft X-rays}

\author{S.Y. Zhou}
\affiliation{Department of Physics, University of California,
Berkeley, CA 94720, USA}
\affiliation{Materials Sciences Division,
Lawrence Berkeley National Laboratory, Berkeley, CA 94720, USA}

\author {\c{C}.\"{O}. Girit}
\affiliation{Department of Physics, University of California,
Berkeley, CA 94720, USA}
\affiliation{Materials Sciences Division,
Lawrence Berkeley National Laboratory, Berkeley, CA 94720, USA}

\author{A. Scholl}
\affiliation{Advanced Light Source, Lawrence Berkeley National Laboratory, Berkeley, California 94720, USA}

\author{C.J. Jozwiak}
\affiliation{Department of Physics, University of California,
Berkeley, CA 94720, USA}
\affiliation{Materials Sciences Division,
Lawrence Berkeley National Laboratory, Berkeley, CA 94720, USA}

\author{D.A. Siegel}
\affiliation{Department of Physics, University of California,
Berkeley, CA 94720, USA}
\affiliation{Materials Sciences Division,
Lawrence Berkeley National Laboratory, Berkeley, CA 94720, USA}

\author{P. Yu}
\affiliation{Department of Physics, University of California,
Berkeley, CA 94720, USA}

\author{J.T. Robinson}
\altaffiliation{Current address: Naval Research Laboratory, Washington, DC 20375}
\affiliation{Materials Sciences Division,
Lawrence Berkeley National Laboratory, Berkeley, CA 94720, USA}
\affiliation{Department of Materials Science and Engineering, University of California,
Berkeley, CA 94720, USA}

\author{F. Wang}
\affiliation{Department of Physics, University of California,
Berkeley, CA 94720, USA}

\author {A. Zettl}
\affiliation{Department of Physics, University of California,
Berkeley, CA 94720, USA}
\affiliation{Materials Sciences Division,
Lawrence Berkeley National Laboratory, Berkeley, CA 94720, USA}

\author{A. Lanzara}
\affiliation{Department of Physics, University of California,
Berkeley, CA 94720, USA}
\affiliation{Materials Sciences Division,
Lawrence Berkeley National Laboratory, Berkeley, CA 94720, USA}

\date{\today}

\begin{abstract}
We study the stability of various kinds of graphene samples under soft X-ray irradiation.  
Our results show that in single layer exfoliated graphene (a closer analogue to two dimensional material), the in-plane carbon-carbon bonds
are unstable under X-ray irradiation, resulting in nanocrystalline structures.  As the interaction along the third dimension increases by increasing the number of graphene layers or through the interaction with the substrate (epitaxial graphene), the effect of X-ray irradiation decreases and eventually 
becomes negligible for graphite and epitaxial graphene.  Our results demonstrate the importance of the
interaction along the third dimension in stabilizing the long range
in-plane carbon-carbon bonding, and suggest the possibility of using
X-ray to pattern graphene nanostructures in exfoliated graphene.
\end{abstract}

\maketitle

The existence and stability of two dimensional materials has
been a fundamental yet long-debated subject. Graphene, a purely one atom thick two dimensional 
material formed by carbon atoms arranged in a honeycomb lattice, was previously 
presumed not to exist because strictly two dimensional materials are
thermodynamically unstable \cite{WMTheorem, Mermin, Landau}.   The recent discovery of graphene \cite{NovoselovSci, PNAS} has raised renewed interest regarding the long debated issue about the stability of two dimensional materials.  Although it has been proposed recently that the ripples - corrugation along the third
dimension in free-standing exfoliated graphene samples help to stabilize graphene \cite{GrapheneRipple, Fasolino, Carlsson}, no direct experimental evidence about what causes the stability of the sp2 bonds has
been obtained so far.  Here we present direct experimental proof that it is indeed the interaction
along the third dimension that drives the stability of the graphene sheet. In particular we show that
the closer the graphene sample is to a two dimenstional crystal, the easier it is for the sp$^2$ bonds to be
broken under soft X-ray irradiation, resulting in nanocrystalline structures. This conclusion was based on a sysmatic study of various kinds of graphene samples
with different amount of interaction along the third dimensionality, including exfoliated graphene on
SiO$_2$, suspended exfoliated graphene, and epitaxial graphene on SiC, and by combining two
important techniques - X-ray absorption spectroscopy which is sensitive to the sp$^2$ bonding, and Raman
spectroscopy which is sensitive to the edges of the graphene samples \cite{Pimenta07, Ferrari2}. Our results show
direct evidence about the instability of two dimensional graphene, and points out a possible route of
using X-rays to engineer graphene nanoribbons.

The samples studied were exfoliated graphene samples on SiO$_2$/Si and epitaxial graphene on
SiC.  The thickness of exfoliated graphene samples is characterized by
optical contrast \cite{ShenNanoLett2007, GeimSci2008} and confirmed by
Raman measurements \cite{Ferrari, Graf}.  The thickness of epitaxial graphene is characterized by low energy
electron microscope (LEEM) \cite{Hibino}.  Although the overall electronic structure of graphene is preserved in both types of graphene samples,
the interaction along the third dimension is much stronger in epitaxial graphene than exfoliated graphene.  This is
manifested by the formation of a first carbon rich layer strongly
bonded with the SiC substrate \cite{Seyller}, and the
opening of a gap in the $\pi$ bands for the first
graphene layer \cite{NatMat, KimAbInitio}.  
X-ray absorption spectroscopy (XAS) spectra at the C 1s edge were
taken with photoemission electron microscope (PEEM) at PEEM2 of the
Advanced Light Source (ALS) in Berkeley.  The microscope is operated
in the total electron yield (TEY) mode by recording the intensity maps
while sweeping the photon energy across the C K edge.  The photon flux
was cut down to only 1/3 of the regular value to decrease the X-ray
radiation damage on the sample.  Raman spectra were taken on a
commercial system (Renishaw inVia) at a wavelength of 514 nm (Argon
laser).  Special care was taken to minimize laser exposure on samples
and subsequent spectra at the same spot showed no change.  The spatial resolution of both techniques allows easy detection of each graphene region with different thickness.

\begin{figure}
\includegraphics[width=6 cm] {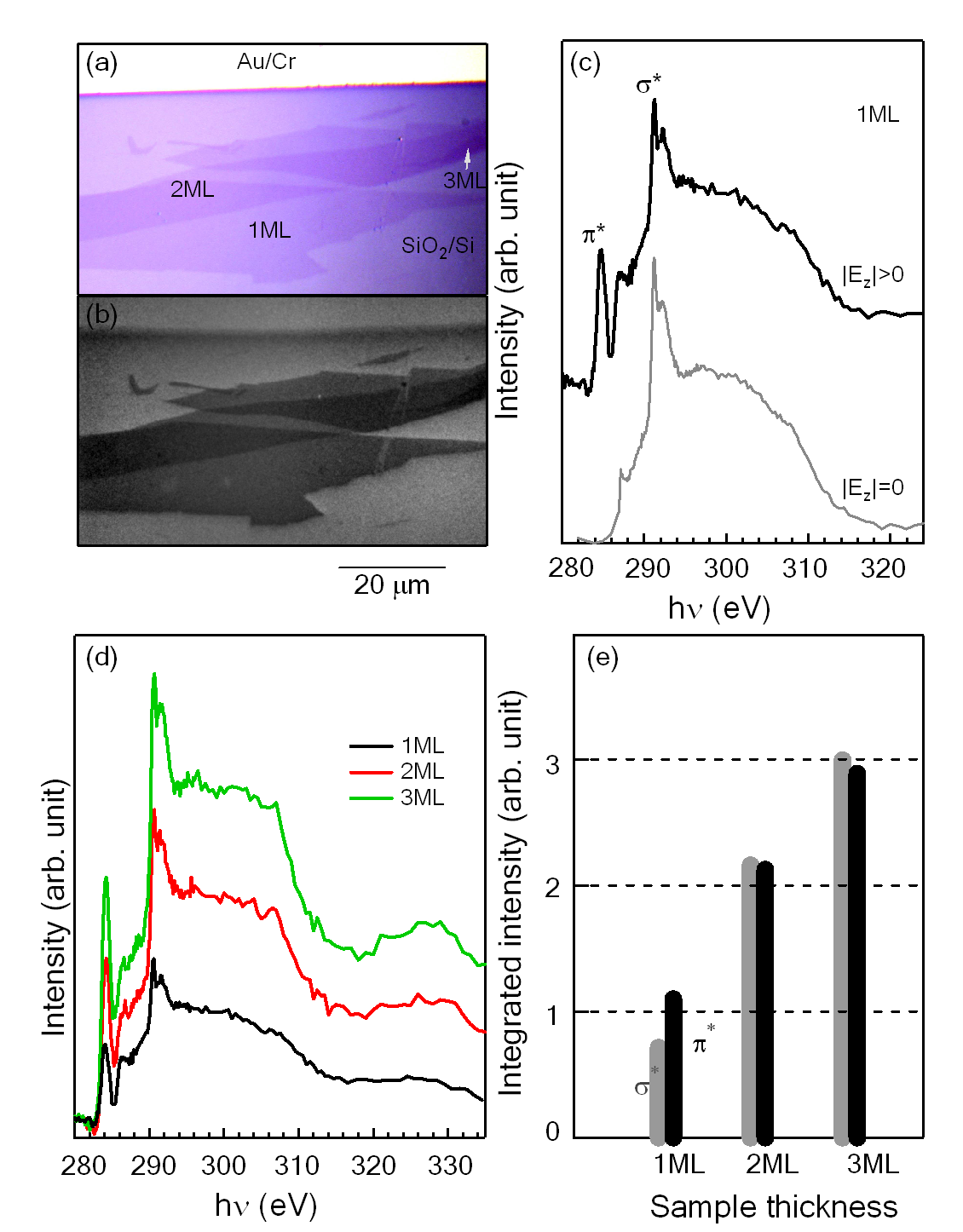}
\label{Figure 1}
\caption{
(Color online) (a) Optical image of a representative exfoliated sample under study with different sample thickness.  The electrical contact to the graphene samples is provided by a region of Au/Cr electrode (light yellow regions on the top  of panel a) deposited on the SiO$_2$.  (b) PEEM image of the same sample taken at photon energy of 279 eV, just below the C 1s absorption edge.  
The spectra are normalized by the spectrum from Au on the same wafer.  (c) C 1s spectra taken on single layer exfoliated graphene with zero (gray curve) and non-zero (black curve) out-of-plane polarization components (E$_z$) respectively. (d) C 1s spectra taken from single layer, bilayer and trilayer graphene with non-zero out-of-plane polarization.  (e) Area of the $\pi^*$ and $\sigma^*$ peaks as a function of the sample thickness extracted from data shown in panel d, after subtraction of a smooth background. The dotted lines are guides for the eye.}
\end{figure}

Figure 1 (a) shows the optical image of a typical exfoliated sample.
Clear optical constrast can be observed between regions of graphene
with different sample thickness \cite{PNAS}.  Figure 1(b) shows the
PEEM image taken at 279 eV photon
energy before the C 1s K edge on the same sample.  
Similar to the optical image (Figure 1a), strong intensity contrasts between different regions are
observed in the PEEM image shown in Figure 1(b).  
Thicker graphene regions will appear darker in the pre-edge spectra because photoelectrons travel through more materials before being emitted to the vacuum, thus making PEEM another powerful tool to capture the thickness of graphene films. 
Figure 1(c) shows the C 1s XAS spectra taken on single layer
exfoliated graphene with two different polarizations.  The two main
features at $\approx$ 285 eV, and at $\approx$ 292 eV, correspond
respectively to the $\pi^*$ orbitals at the K and M points of the
Brillouin zone and to the $\sigma^*$ orbitals at the $\Gamma$
point \cite{PolarizationDep}. Their in-plane and out-of-plane
characters are confirmed by the dependence from the polarization of the
incident photons (Figure 1(c)) \cite{PolarizationDep}.  More specifically,
when the light polarization is in-plane, only the in-plane $\sigma^*$
orbital at 292 eV contributes to the C 1s edge, while when the
out-of-plane polarization component increases, the intensity of the
$\pi^*$ feature at 285 eV strongly increases.

In Figure 1(d) we show the C 1s spectra of exfoliated graphene for
different sample thickness.  As in Figure 1(c), the $\pi^*$ feature and
the splitting of the $\sigma^*$ orbitals are observed in all the
spectra, independent of the sample thickness.  The main difference is
that as the thickness increases, the intensity of the $\sigma^*$ and
$\pi^*$ features increases. 

The high energy resolution and sample quality of this experiment has enabled us to clearly
resolve the splitting of the $\sigma^*$ peaks, not observed in a previous
study \cite{GrioniPEEM}, and hence to quantify the thickness
dependence of the absorption spectra and directly compare it to the
optical reflectivity data.  In Figure 1(e) we show the integrated absorption
intensity, determined from the area underneath the $\pi^*$ and
$\sigma^*$ orbitals near the C K edge.  The data clearly show that the
intensity of the $\pi^*$ and the $\sigma^*$ orbitals scales linearly
with the sample thickness for up to three layers.  Interestingly, this
is in close analogy to the quantized steps observed in optical
reflectivity and transmission for different graphene
thickness \cite{ShenNanoLett2007, GeimSci2008}, which are defined only
by the fine structure universal constant $\alpha=e^2/\hbar
c$ \cite{GeimSci2008}.  
This result suggests that, even at higher photon energies the universality still holds and the absorption spectra is
still related to the universal constant $\alpha$, the parameter that
defines the optical transparency of graphene.

\begin{figure*}
\includegraphics[width=12.8 cm] {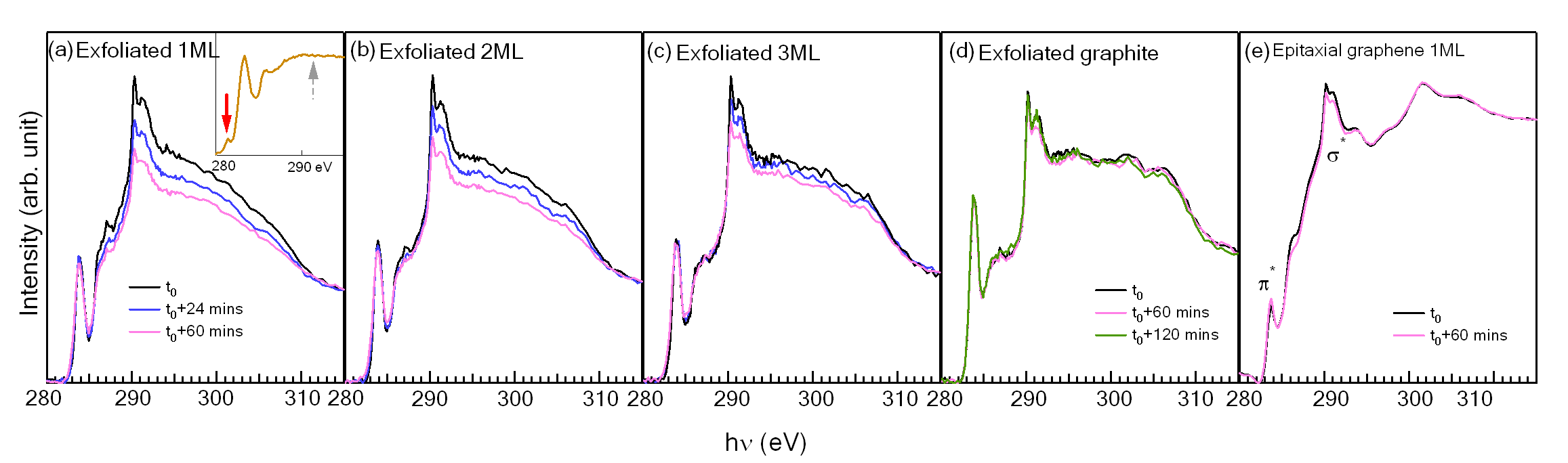}
\label{Figure 2}
\caption{
(Color online) Time dependence of the C 1s XAS for (a) single layer (b) bilayer and (c) trilayer exfoliated graphene (d) exfoliated graphene and (e) epitaxial single layer graphene.  The spectra are offset to have zero intensity in the pre-edge and normalized to the same value at 315 eV for each sample.  The inset in panel a shows an extreme sample which has negligible $\sigma^*$ signal (pointed to by the gray arrow) after irradiation and additional pre-edge peak (pointed to by the red arrow) which is likely asscoiated with edge states.}
\end{figure*}

Figure 2 shows the time evolution of the absorption spectra for
exfoliated (panels a-d) and epitaxial (panel e) graphene.  The
relative intensity ratio of the $\sigma^*$ to the $\pi^*$ orbital show
a clear time dependence from X-ray exposure. More specifically, the
time dependence of the spectra is strongest for exfoliated single layer
graphene (Figure 2(a)), as measured from the 24$\%$ change in the
intensity ratio of the $\sigma^*$ orbital to the $\pi^*$ orbital, and
decreases significantly as the number of layer increases, from 20$\%$
in bilayer to 15$\%$ in trilayer graphene.  These findings suggest a
strong radiation-induced damage of the exfoliated graphene upon soft
X-ray exposure, a surprising result considering the inert character of carbon.
This is reminiscent of the radiation-induced damage observed in carbon rich biological samples \cite{Molodtsov}.  We note that, although the radiation-induced damage presented here is observed for all the
samples studied (more than five samples in total), in some cases the absorption spectra
of single layer graphene shows more dramatic changes.  In an extreme
case (inset in Figure 2(a)), the signal of the $\sigma^*$ orbital
completely disappeared (pointed to by a gray arrow) and an additional peak arose before the
adsorption edge (pointed to by a red arrow), reminiscent of the edge
states observed in nanographite sample \cite{EdgeStates}.

The strong radiation-induced change of the single layer exfoliated
graphene (Figure 2(a)) is in contrast to the almost negligible effect observed on the
epitaxial sample (Figure 2(e)).
As for the exfoliated sample, the two main features in the spectra are the $\pi^*$ and the $\sigma^*$ orbitals, which occur at the same energy, a further confirmation
that the overall electronic structure of exfoliated and epitaxial graphene is similar in nature.
In this case however, the relative intensity of the $\pi^*$ to the $\sigma^*$ peak is much weaker, suggesting that the $\pi^*$ derived states are partially occupied.  This is likely due to a charge transfer from the buffer layer, in agreement with photoemission measurements reporting an highly doped nature of this sample \cite{NatMat}.
Note that, in addition to these main features, other peaks are observed above the $\sigma^*$ peak, likely due to the substrate, as observed from a direct comparison with the absorption spectra of SiC \cite{SiCXAS}.

The time dependence of the spectra in Figure 2(e) clearly shows that the epitaxial sample is very stable and the damage induced by x-ray irradiation is negligible, despite the much longer exposure time.  This result is in striking contrast with the strong time dependence observed in the exfoliated sample (Figure 2a).  Note that these two types of samples were measured in the same chamber under the same conditions, and therefore this substantial difference in stability should be attributed to their intrinsic nature, e.g. the different strength of interaction along the third dimension, and not an artifact of the measurements.  

\begin{figure}
\includegraphics[width=6. cm] {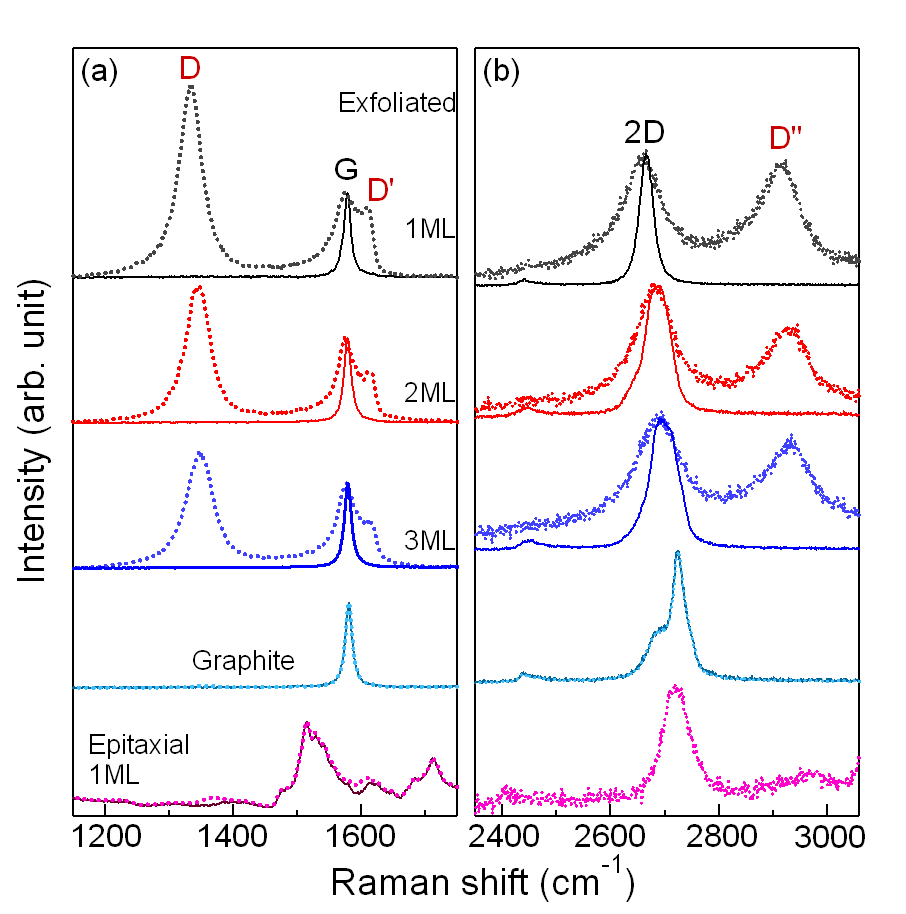}
\label{RamanfJune27}
\caption{
(Color online) Raman spectra for exfoliated samples with different sample thickness (1ML, 2ML, 3ML and graphite) and single layer epitaxial graphene with laser excitation wavelength of 514 nm in the energy range near the D (a) and 2D peak (b).  The solid and the dotted curves are taken before and after X-ray irradiation.  The spectra are normalized to have the same amplitude for the G peak and 2D peak.}
\end{figure}

To better understand the origin of this radiation-induced change, we
compare Raman measurements on the same samples measured before and after
exposure to soft X-rays (see Figure 3).  Before
exposure, the exfoliated graphene sample shows a sharp G peak induced by the zone center E$_{2g}$ phonon at $\approx$ 1570 cm$^{-1}$, and a
strong 2D peak (overtone of the zone boundary A$_{1g}$ phonon) at
$\approx$ 2700 cm$^{-1}$.  The 2D peak in single layer graphene shows
a single component, while the 2D peak for bilayer, trilayer graphene and
many layer graphite is broader and contains multiple components,
in agreement with previous studies \cite{Ferrari, Graf}.  The
sharpness of the G and 2D peaks, as well as the absence of the
disorder-induced D peak at $\approx$ 1360 cm$^{-1}$ \cite{Pimenta07,
Ferrari2, Vidano}, 
D$^{\prime\prime}$ peak at $\approx$ 2950 cm$^{-1}$, shows that the
samples under study are of very high quality, and the amount of
disorder is negligible.  
However, after X-ray exposure, the Raman
spectra show significant changes in all the exfoliated samples,
although the degree of change decreases from single layer to trilayer
graphene and almost disappear for graphite, in line with the absorption results previously discussed. 
More specifically, we observe the appearence of a huge D peak at 1360
cm$^{-1}$, as well as the appearance of D$^\prime$ peak at 1620
cm$^{-1}$ and D$^{\prime\prime}$ peak at $\approx$ 2950 cm$^{-1}$.  These changes are
accompanied by the broadening of both the G (1570 cm$^{-1}$) and 2D
(2700 cm$^{-1}$) peaks.  In line with the absorption results (see Figure 2), the comparison between graphene samples with different thickness
suggests that the largest radiation-induced changes occur on the
single layer exfoliated graphene sample.  On the contrary, the Raman
spectrum for the single layer epitaxial graphene shows negligible
change, although it has been exposed to X-ray for a much longer time
($\approx$ 10 hours), once again pointing to the different nature
between exfoliated and epitaxial graphene, and a better stability of
the latter to X-ray irradiation.


The results presented here point to a scenario where the X-rays locally break the sp$^2$ bonding 
giving rise to small crystallites within the same graphene sheet.  
This is supported by: 1) the decrease, or in some cases even the disappearence, of the $\sigma^*$-orbital-related signal in the absorption spectra (Figure 2), suggesting a decrease of sp$^2$ bonded carbon atoms; 
2) the appearence of a broad pre-edge feature (inset of figure 2(a)) in the absorption spectra, similar to that observed in nanographite edges \cite{EdgeStates}; 3) the appearence of strong D, D$^\prime$ peaks in the Raman spectra (Figure 3), associated with an increase of broken carbon bonds and edges states
\cite{Pimenta07, Ferrari2, Vidano, Graf, FerrariRamanDisorder}; 4) the broadening of the G and 2D peak in the Raman spectra, suggestive of phonon confinement.  The average crystallite size estimated from the ratio between the G and D peaks in the Raman
spectra \cite{Pimenta07} changes from 7 nm in single layer exfoliated graphene to 10 and 12 nm in bilayer and trilayer exfoliated graphene respetively, pointing to a smaller irradiation induced bond-breaking effect in thicker exfoliated graphene samples.

There are a few possibilities about the bond-breaking mechanism, and further investigation is needed to pin down the exact mechanism.  First, the higher probability of bond breaking by X-rays in insulating samples \cite{BondbreakingMechanism} 
suggests that the bond-breaking in graphene, which is a semimetal, might be caused by the ejection of electrons into the vacuum and the insufficient restoration of the ejected electrons as a result of its low charge carrer concentration.  This is particularly so considering that the SiO$_2$ layer underneath the graphene sample is quite insulating as well.  
Another possibility is that in exfoliated graphene, the SiO$_2$ substrate or even residual water between graphene
and the SiO$_2$ layer might act as a {\it``catalyst''} for the sp$^2$ bond-breaking 
through formation of epoxy groups or carbon mono-oxide groups.  
To test the second possiblity, we have measured two bilayer graphene samples partially suspended on a few $\mu$m width trenches and we have detected a very similar amount of D peak in both the suspended and unsuspended regions after X-ray irradiation.  This suggests that whatever the bond breaking mechanism is, it must be intrinsic to the exfoliated graphene, and irrelavant of the SiO$_2$ substrates or the trapped water or gas between graphene and the SiO$_2$.  As shown above, the  radiation-induced bond-breaking effect also decreases
when the interaction along the third dimension increases either by
increasing the number of layers or through a
stronger interaction with the substrate, suggesting the important role of the interaction along the third dimensionality.

In summary, we found that single layer exfoliated graphene is unstable
under soft X-ray exposure, resulting in local breaking of the sp$^2$
bonding and the formation of small crystallites within the sample.  As
the interaction along the third dimension increases, either by
increasing the sample thickness or through stronger bonding with the
substrate, as in the case of epitaxial graphene, the graphene sample
becomes more stable.  These results point to the crucial role of the
interaction along the third dimension in stabilizing quasi-two
dimensional graphene.  Finally, this ability to easily break the
carbon-carbon bonds suggests the possiblity of using X-rays or even
lasers \cite{LaserRadiation} to ``write'' graphene nanostructures, an alternative to standard writing techniques.  Further studies are needed to investigate where the bonds break \cite{CogSci} and to achieve a fine control of the amount of broken bonds.


\begin{acknowledgments}
We thank D.-H. Lee, A.K. Geim and A.C. Ferrari for useful discussions. The photoemission and Raman measurements were supported by the Division of Materials Sciences and Engineering, Office of Basic Energy Sciences of the U.S. Department of Energy under Contract No. DE-AC03-76SF00098.  AZ and COG acknowledge the DOE grant DE-AC02-05CH11231 for sample preparation and Raman characterization.
\end{acknowledgments}

\end{document}